\documentclass[aps,twocolumn,showpacs,floatfix]{revtex4-1}

\usepackage[caption=false]{subfig}
\usepackage{xcolor}
\usepackage{verbatim}
\usepackage{graphicx}
\usepackage{dcolumn}
\usepackage{bm}
\usepackage{color}
\usepackage[colorlinks=true, allcolors=blue]{hyperref}
\usepackage{makeidx}
\usepackage{amsmath}
\makeindex
\def\>{\right\rangle}
\def\<{\left\langle}
\def\be{\begin{equation}}
\def\ee{\end{equation}}
\def\ba{\begin{aofprray}{l}}
\def\ea{\end{aofprray}}

\def\beq{\begin{eqnarray}}
\def\eeq{\end{eqnarray}}

\begin{document}
\title{Non-equilibrium effects on charge and energy partitioning after an interaction quench}
\author{Alessio Calzona$^{1,2,3}$, Filippo Maria Gambetta$^{1,2}$, Matteo Carrega$^4$, Fabio Cavaliere$^{1,2}$, and Maura Sassetti$^{1,2}$}
\affiliation{ $^1$ Dipartimento di Fisica, Universit\`a di Genova, Via Dodecaneso 33, 16146, Genova, Italy.\\
$^2$ SPIN-CNR, Via Dodecaneso 33, 16146, Genova, Italy.\\
$^3$ Physics and Materials Science Research Unit, University of Luxembourg, L-1511 Luxembourg. \\ 
$^4$ NEST, Istituto Nanoscienze-CNR and Scuola Normale Superiore, I-56127 Pisa, Italy.
} 
\date{\today}
\begin{abstract}
Charge and energy fractionalization are among the most intriguing features of interacting one-dimensional fermion systems. In this work we determine how these phenomena are modified in the presence of an interaction quench. Charge and energy are injected into the system suddenly after the quench, by means of  tunneling processes with a non-interacting one-dimensional probe. Here, we demonstrate that the system settles to a steady state in which the charge fractionalization ratio is unaffected by the pre-quenched parameters. On the contrary, due to the post-quench non-equilibrium spectral function, the energy partitioning ratio is strongly modified, reaching values larger than one. This is a peculiar feature of the non-equilibrium dynamics of the quench process and it is in sharp contrast with the non-quenched case, where the ratio is bounded by one. 
 
\end{abstract}
\maketitle

\section{Introduction}

One-dimensional (1D) systems are an ideal playground where both non-equilibrium phenomena and interaction effects can be studied. Many peculiar behaviors have been predicted to occur in 1D~\cite{Haldane:1981,giamarchi2003quantum,Voit:1995}, with a richer and more complicated physics compared to analogous case in higher dimension. For instances, it is well-known that fermionic particles in 2 or 3 dimensions are well-described by the Fermi liquid theory, where interactions do not play a dramatic role~\cite{Landau,Giuliani:2010}. Indeed, for these systems one can re-formulate the problem in terms of quasiparticles with parameters renormalized by interactions, such as the effective mass or the particle velocity.
On the contrary, in 1D the Fermi liquid paradigm fails and interactions crucially influence the whole dynamics of the system itself~\cite{Haldane:1981,giamarchi2003quantum,Voit:1995}.
Thanks to the great advances in nanotechnology, experimental realizations of 1D and quasi-1D systems recently have lead to the verification of different predictions and to the observation of intriguing phenomena. 
Few examples of condensed matter systems which exhibit typical 1D behavior include edge states in topological materials~\cite{vonKlitzing:1986,Stormer:1999,Wu:2006,Konig:2007,Carrega:2011,Wang:2015,Dolcetto:2016}, carbon nanotubes~\cite{Bockrath:1999,Egger:1997,Ishii:2003}, semiconducting nanowires~\cite{auslaender01042005,Jompol:2009,Quay:2010,SOC}, atomic chains~\cite{Blumenstein:2011} and Bechgaard salts~\cite{Vescoli:2000}.

Among all the peculiar features of interacting 1D gapless fermion systems,   charge~\cite{maslov1995landauer,safi1995transport,safi1997properties,steinberg2007charge,deshpande2010electron,kamata2014fractionalized,Trauzettel:2004,Milletari:2013,Inoue:2014,Schneider:2016,bena2001measuring,Wahl:2014} and energy fractionalization~\cite{Karzig:2011,calzona2016extended} are some of the most interesting ones. Here, when a particle is injected into an interacting 1D system it splits up and originates two collective excitations that propagate in opposite directions, each one carrying a fraction of the particle original charge and energy.
The partitioning of these quantities between the two excitations depends on the strength of the interparticle interaction and, in the case of the energy, also on the injection process~\cite{Karzig:2011, calzona2016extended}. In the context of fractionalization phenomena many theoretical predictions have been put forward~\cite{pham2000fractional,le2008charge,das2011frac,Karzig:2011,perfetto2014time,calzona2015spin,calzona2015transient,calzona2016extended} and recently charge fractionalization has been experimentally tested in different 1D interacting systems~\cite{steinberg2007charge,barak2010interacting,kamata2014fractionalized,Inoue:2014}. 

From a theoretical point of view, the low energy sector of 1D interacting systems belong to the universality class of  the integrable Luttinger Liquid (LL)~\cite{Voit:1995,vondelft,Haldane:1981,giamarchi2003quantum,miranda}.
LL theory is a powerful tool for the study of standard equilibrium properties, and also allows for the investigation of out-of-equilibrium physics \cite{Cazalilla:2006,Uhrig:2009,Iucci:2009,Cazalilla:2016} of interacting systems.
This latter topic was recently addressed and investigated especially in view of the experimental progresses in ultracold atomic gases~\cite{Bloch:2008rmp,Bloch:2008s,Bloch:2012} which has renewed the interest in this field~\cite{Polkovnikov:2011,Eisert:2015}. Indeed, the possibility of tuning with high precision and in a time-dependent fashion some of the system parameters, such as the interaction strength~\cite{Bloch:2008rmp,Langen:2015,Loftus:2002,Roberts:1998,Greiner:2002,Greiner:2002b}, allowed to probe their real time evolution and to perform transport
experiments~\cite{Kinoshita:2006,Brantut:2012,Krinner:2015,Husmann:2015,Krinner:2015b,Weitenberg:2011,Brantut:2012,Trotzky:2012,Cheneau:2012,Fukuhara:2013,Krinner:2015,Krinner:2015b,Husmann:2015}.
Recently, a different approach to inspect  non-equilibrium effects in 1D system has been developed in the context of quantum Hall edge states. Here, a non-equilibrium energy distribution can be obtained by studying a quantum point contact connecting two edge states with different chemical potential~\cite{ExpAltimiras:2010,Degiovanni:2010,Kovrizhin:2011,Levkivskyi:2012,Cazalilla:2016,Milletari:2013,Inoue:2014,Schneider:2016}.
One natural question about isolated interacting quantum many body systems far from equilibrium is  whether they thermalize or not. It has been shown that this is indeed the case for the vast majority of quantum systems~\cite{vonNeumann:1929,Goldstein:2010, Polkovnikov:2011,Eisert:2015}. However, if a system is integrable~\cite{Caux:2011}, as it is for the LL, it can relax to a stationary non-thermal state, retaining strong memory of its initial conditions~\cite{Rigol:2007,Rigol:2008,Cassidy:2011,Kim:2016,Yin:2016}.
Recent works have indeed confirmed how the equilibrium spectral and transport properties of  a LL~\cite{Kane:1992,Matveev:1993,Voit:1993,Bockrath:1999,Furusaki:2002} are modified by an interaction quantum quench~\cite{Kennes:2013,Kennes:2014,Perfetto:2010,Schiro:2014,Schiro:2015,gambetta2016transient,Porta:2016}. 
In this respect a still open question is how fractionalization will be influenced by quench and how strong are the memory effects of the initial pre-quenched state. 

This issue will be addressed in the present paper. We consider a 1D system subjected to a sudden quench of the interparticle interaction and non-local tunnel coupled to a parallel voltage biased 1D non-interacting probe.
We demonstrate that in the long time limit the system settles to a steady state in which the ratio of the charges that travel along the two directions depends only on the post-quench Hamiltonian without any memory of the initial state. This is a quite remarkable behavior, since the majority of the observables of a quenched integrable model retain a strong memory of the initial state~\cite{Polkovnikov:2011,Cazalilla:2016}. The reason for that can be traced back to the absence of charge transfer between the LL channels.  On the contrary, the energy partitioning is strongly affected by the quench with a finite memory of the initial state and sharp differences from a non-quenched case.  We explain this behavior as induced by the modifications in the spectral function of a quenched LL~\cite{Kennes:2014}.  Indeed, due to the excited nature of the steady state, the spectral function possesses non vanishing tails which allow for energy flow in energy-momentum regions not accessible in the non-quenched case. As a consequence, the differential energy current in one of two directions can be negative and the energy partitioning ratio can thus be greater than one.

The paper is organized as follows. In Sec.~\ref{sec:Model} we develop the model for the two voltage biased parallel 1D fermion systems, describing the interaction quench and the non-local tunneling. The analysis of the charge fractionalization in the quenched LL and its independence from the initial state is described in Sec.~\ref{sec:chargefrac}, while the behavior of energy partitioning and its dependence on quench is discussed in Sec.~\ref{sec:energypart}. Sec.~\ref{sec:conclusions} contains the summary of our results. 

\section{Model and general setting}\label{sec:Model}
\begin{figure}
	\centering
	\includegraphics[width=\linewidth]{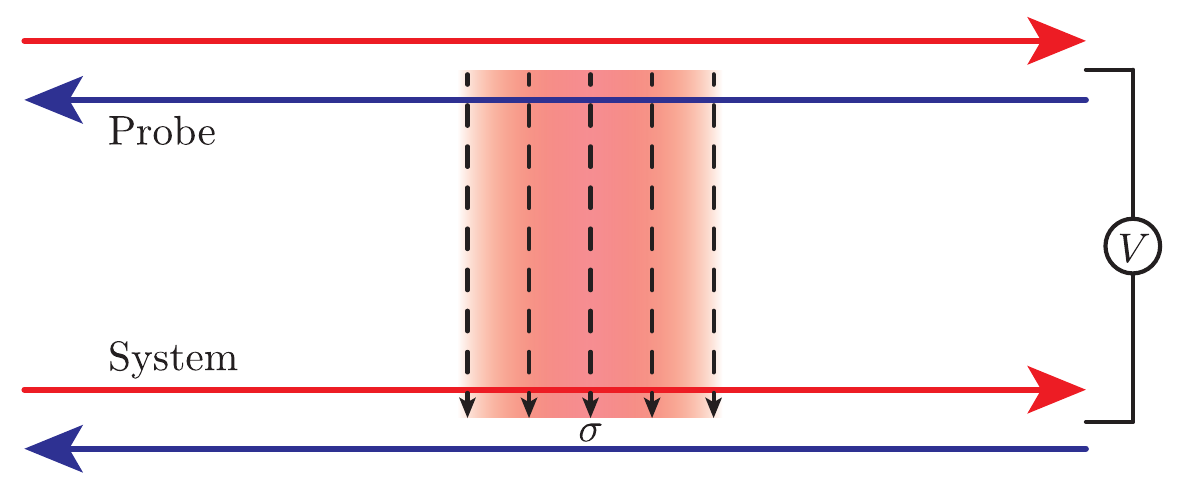}
	\caption{(Color online) Schematic representation of the two parallel 1D fermion systems. The probe and the system, modeled as a pair of counterpropagating channels, are tunnel-coupled over a region of finite size $ \sigma $ and biased with a dc voltage $ V $. In order to study fractionalization phenomena, we consider only the injection from the $ L $-channel of the probe to the $ R $-channel of the system, as highlighted by the shaded red region. See text for further details.  } 
	\label{fig:setup}
\end{figure}

We consider two parallel 1D fermion systems: one which play the role of the external probe treated as a non-interacting LL (hereafter referred as the probe) and the other representing the interacting LL (called the system) which undergoes a quantum quench of the interparticle  interactions. A schematic view of the setup is depicted in Fig.~\ref{fig:setup}. The system and the probe are modeled in terms of a pair of counterpropagating channels, denoted by $r=R,L$ depending on their right- or left-moving nature \cite{nota1}. 
The probe is described by a non-interacting linearized Hamiltonian (throughout this paper, we set $\hbar = 1$)
\begin{equation}
\hat H_p = v_F  \sum_{r=R,L} \vartheta_r \int_{-\infty}^{+\infty} dx\; \hat \chi_r^\dagger(x) (-i  \partial_x) \hat \chi_r(x),
\end{equation}
where $v_F$ is the Fermi velocity, $\vartheta_{R/L} = \pm1$ and $\hat \chi^\dagger_r(x)$ is the fermion field associated to the $ r $-channel.\\
The system Hamiltonian $\hat H_s $ contains a free part and an interacting one, { i.e.} 
\begin{equation}
\hat H_s = v_F \sum_{r=R,L}  \vartheta_r \int_{-\infty}^{+\infty} dx\; \hat \psi_r^\dagger(x) (-i \partial_x) \hat \psi_r(x) + \hat H_{int},
\label{Hsystem}
\end{equation}
where $\hat \psi_r^\dagger(x)$ is the fermion field of the $r$-channel and 
\begin{equation}
\hat H_{int}= \frac{g_4}{2} \sum_r \! \int_{-\infty}^{+\infty} \! dx \left[\hat n_r(x)\right]^2 + g_2 \int_{-\infty}^{+\infty}\! dx \; \hat n_R(x) \hat n_L(x)
\end{equation}
describes interactions in the system.
Here, $\hat n_r = \,:\!\hat \psi_r^\dagger \hat \psi_r\!\!: $ is the particle density on the $r$-channel and $g_4$, $g_2$
model the interaction strengths in the LL language, referring to the intra- and inter-channel interactions respectively \cite{vondelft,Voit:1995,giamarchi2003quantum,miranda}.
At time $t=0$ we assume that a sudden quench of the interaction occurs. Therefore the latter parameters will change in time following the behavior $g_{2,4}(t) =  g_{2,4}^{i}\,\theta(-t) + g_{2,4}^{f}\,\theta(t)$~\cite{Cazalilla:2006,Cazalilla:2016,gambetta2016transient}.
Before the quench protocol, i.e. for $t<0$, both the probe and the system are prepared in their own ground state dictated by the previous Hamiltonian.

Non-local tunneling between probe and system is switched on immediately after the quench. To this end, the probe is subjected to a bias voltage energy $eV$, measured with respect to the Fermi level of the interacting system.
Since our goal is to study fractionalization phenomena it is necessary to break the inversion symmetry injecting only either $R$- or $L$-moving particles (see the shaded red region in Fig.~\ref{fig:setup}). This can be achieved, for instance, via momentum resolved tunneling \cite{auslaender01042005,steinberg2007charge,barak2010interacting}, a technique that played a crucial role in the first experimental detection of charge fractionalization in 1D quantum wires \cite{steinberg2007charge}. It consists in using an external magnetic field $\mathbf{B}$, perpendicular to the tunneling plane, that effectively shifts in momentum the dispersion relation of the probe by an amount  $|\delta k| = e B d$ ($d$ is the distance between the probe and the system). Such a shift can be tuned so that the tunneling can occur only between, say, the probe $L$-channel and the system $R$-channel. In that case the non-local tunneling Hamiltonian reads \cite{dolcetto12extended,chevallier10extended, vannucci}
\begin{equation}
\label{eq:tunneling}
\hat H_t = \hat H_t^++\hat H_t^-(t) = \theta(t) \int_{-\infty}^{+\infty} \!\!dx \, \xi(x)
\;  \hat \psi_R^\dagger(x) \hat \chi_L(x) + \text{h.c.}.
\end{equation}
The tunneling amplitude function $\xi(x) = \lambda e^{-x^2/\sigma^2}(\sigma \sqrt{\pi})^{-1}$ models the extended tunneling region with width $\sigma$, while the Heaviside function takes into account the switching on of the tunneling at $t=0^+$. Since the probe is also subjected to a bias voltage $V$ it is worth to  consider two parameters: $k_P$, the momentum of a probe $L$-particle at the Fermi level and $eV$, the energy of the most energetic particle in the probe. Then, the momentum $k_0 = k_P - eV/ v_{{F}} - k_F$ will represent the momentum difference between the most energetic $L$-particle of the probe and the Fermi momentum $k_F$ of the system $R$-channel.
 
We conclude this part  by recalling the standard procedure of interaction diagonalization in the presence of a sudden quench.
Bosonization technique \cite{vondelft,miranda,Haldane:1981,giamarchi2003quantum,Voit:1995} allows to represent the fermion fields  $\hat \psi_r(x)$ in terms of free boson fields $\hat \phi_r(x)$ (the ones which diagonalize the fermionic free part in Eq.(\ref{Hsystem}))
\begin{equation}
\label{eq:boson_eq}
\hat \psi_r^\dagger (x) = \frac{1}{\sqrt{2 \pi a}}  e^{i \sqrt{2 \pi} \hat \phi_r(x)} \; e^{-i \vartheta_r k_F x}.
\end{equation}
Here, so-called Klein factors have been safely omitted and we have denoted with $a$ the cut-off length. Within bosonization the system Hamiltonian density can be diagonalized. In the post-quench regime, i.e. for $t>0$, one has 
\begin{equation}
\mathcal{\hat H}_s (x,t) = \frac{u}{2} \sum_{\eta=\pm} \left[\partial_x \hat \phi_\eta(x-\eta u t)\right]^2 ,
\end{equation}
where $u=(2\pi)^{-1}\sqrt{\big(2\pi v_F+g_4^{f}\big)^2-\big(g_2^{f}\big)^2}$ is the propagation velocity, renormalized by interactions. The boson fields $\hat \phi_\eta(x)$ are chiral, with $\eta=\pm$ referring to the propagation direction. They are related to the free boson fields $\hat \phi_r(x)$ in Eq. \eqref{eq:boson_eq} by means of  the linear combination
\begin{equation}
\hat \phi_r(x) = \sum_{\eta=\pm} A_{\eta \vartheta_r} \hat \phi_\eta(x),\qquad
2 A_\pm = \frac{1}{\sqrt{K^f}}\pm\sqrt{K^f},
\end{equation}
where we have introduced the dimensionless interaction parameter~\cite{vondelft,miranda,giamarchi2003quantum,Voit:1995}
\begin{equation}
K^\mu= \sqrt{\frac{2 \pi v_F +g_4^{\mu} -g_2^{\mu}}{2 \pi v_F +g_4^{\mu} +g_2^{\mu}}}\qquad\text{with }\mu=\{i,f\}.
\end{equation} 
We recall that for a fermionic system with repulsive interactions $K^\mu$ can take values $0< K^\mu \leq 1$, with $K^\mu=1$ representing the non-interacting limit.
The corresponding chiral particle density operators are~\cite{vondelft,miranda,giamarchi2003quantum,Voit:1995}
\begin{equation}
{\hat  n_\eta(x-\eta u t) = -\eta\, \sqrt{\frac{K^f}{2 \pi}} \; \partial_x \hat \phi_\eta(x-\eta u t)}.
\end{equation}
Before the quench ($t<0$), the interaction parameter $K^i$ is different from $K^f$ and the pre-quench chiral diagonal fields $\hat \varphi_\eta(x)$ are related to the post-quench ones by 
$\hat \phi_\eta(x)= \sum_{\nu=\pm} \alpha_{\nu \eta} \hat \varphi_\nu(x)$ \cite{gambetta2016transient}. Here, factors  
\begin{equation}
 2 \alpha_\pm= \sqrt{\frac{K^i}{K^f}} \pm \sqrt{\frac{K^f}{K^i}}
\end{equation}
are sensitive to the initial and final interaction strength.
Finally, we remind the reader that the LL is a low energy effective theory which relies on the linearization of energy dispersion relation around the Fermi energy. Thus, the range of momenta and bias voltages that can be addressed by our analysis is restricted to a region of the $ (V,k)-$space in which the bands of the physical system can be well described as linear ones~\cite{Imambekov:2012}. This is usually true for energies smaller than the Fermi energy of the system. However, for Dirac materials, such as 2D topological insulators, the linear region of the spectrum extends over the entire gap~\cite{Dolcetto:2016}.

\section{Interaction quench and charge fractionalization}\label{sec:chargefrac}
In order to study the fractionalization of the injected charge current, we compute the total amount of injected charge that travels in the $\eta$ direction along the system. Its time derivative is directly the charge chiral current, and in the steady state limit is given by \cite{nota2}
\begin{equation}
\label{eq:IQ}
I^Q_\eta = \lim_{t \to \infty} \partial_t \int_{-\infty}^{+\infty} dx\; \langle\delta n_\eta(x,t)\rangle,
\end{equation}
where
\begin{equation}
\label{eq:av}
 \langle\delta n_\eta(x,t)\rangle = \text{Tr} \left\{\hat n_\eta(x,t) \left[\hat \rho (t)-\hat \rho(0)\right]\right\}
\end{equation}
is the average variation of the chiral particle number induced by tunneling and $ \hat{\rho}(t) $ the time-dependent total density matrix. The time evolution in the trace is evaluated in the interaction picture with respect to $ \hat{H}_t $. Here, we assume that the total system is in thermal equilibrium immediately before the quench and $ \hat{\rho}(0) $ is the associated equilibrium density matrix. The average in Eq.~\eqref{eq:av} is performed in Appendix \ref{app:charge} following the lines described in details in Appendix~\ref{app:operatorO}, where the average variation of a generic hermitian and particle number conserving operator $\hat O(x,t)$ is computed. At the lowest order in the tunneling amplitude it reads
\begin{equation}
\label{eq:O_general}
\begin{split}
&\langle\delta  O (x,t)\rangle = 2 \Re \int_{0}^{t}\!d\tau_1\!\!\int_{0}^{\tau_1}\!d\tau_2 \!\! \int_{-\infty}^{+\infty}\! dy_1 dy_2\; \xi^*(y_2)\xi(y_1) \\ &\!\times \left\{\langle{\hat \chi_L^\dagger(y_2,\tau_2) \hat \chi_L(y_1,\tau_1)}\rangle \langle{\hat \psi_R(y_2, \tau_2) \left[\hat O(x,t), \hat \psi_R^\dagger (y_1,\tau_1)\right]}\rangle \right. \\  &\!+ \left. \langle{\hat \chi_L(y_2,\tau_2) \hat \chi_L^\dagger(y_1,\tau_1)}\rangle \langle{\hat \psi_R^\dagger(y_2, \tau_2) \left[\hat O(x,t), \hat \psi_R (y_1,\tau_1)\right]}\rangle \right\}.
\end{split}
\end{equation}
Here the notation $\langle \dots \rangle$ stands for the quantum average over the generic pre-quench states of both the system and the probe. From now on we focus on the $T=0$ limit, assuming that before the quench both the system and the probe are in their ground state. Eq. \eqref{eq:O_general} with $\hat O(x,t)= \hat n_\eta(x,t)$ allows us to obtain the chiral charge current given in Eq. \eqref{eq:IQ_app}. 
	
The contribution of particles with well-defined energy $eV$ to the charge current can be identified by focusing on the {\it differential} chiral charge current
\begin{equation}
\label{eq:G_Q}
\begin{split}
\mathcal{G}^Q_\eta = \frac{\partial I^Q_\eta}{\partial V} \Bigg|_{k_P},
\end{split}
\end{equation}
which has to be evaluated keeping the probe dispersion relation fixed, i.e. with $k_P$ constant. As it is well known, the differential charge current is intimately related to the spectral function of the system~\cite{Bruus:2004}. 
Since we are studying the injection of $R$-particles in the system, we focus on its $R$-branch spectral function which is defined as 
\begin{equation}
\mathcal{A}(\epsilon,k) = \int_{-\infty}^{\infty}\!\! d\tau dx\; e^{i \epsilon \tau - i k x} \lim_{t \to \infty} \mathcal{A}(x,\tau,t),
\end{equation}
where
\begin{equation}
\mathcal{A}(x,\tau,t) = \Big\langle \left\{\hat \psi_R(x,t+\tau) , \hat \psi_R^\dagger(0,t)\right\}\Big\rangle.
\end{equation}
The average is evaluated with respect to the pre-quench ground state. This relation is, for instance, at the basis of the celebrated tunneling spectroscopy technique~\cite{Bruus:2004}. In Appendix \ref{app:spectral} we thus show that 
\begin{equation}
\label{eq:GQeta}
\mathcal{G}^Q_\eta =  \frac{e\sqrt{K^f} A_\eta}{4 \pi^2 v_{F}} \int_{-\infty}^{\infty}\,\, dk\; \left|\tilde \xi (k-k_0)\right|^2 \mathcal{A}(eV,k),
\end{equation} 
where $\tilde \xi(k) = \lambda \exp(-k^2\sigma^2/4)$ is the Fourier transform of the function $\xi(x)$.\\
 This result can be understood considering, at first, a tunneling region with infinite width, i.e. $\sigma \to \infty$. In this limit, the momentum $k_0$ is conserved through the whole tunneling process, as well as the particles energy $eV$. Thus, one can expect $ \mathcal{G}^Q_{\eta}\propto\mathcal{A} $, with $ \mathcal{A}(eV,k_0) $ representing the probability of finding a system excitation with energy $eV$ and momentum $k_0$. When a finite width $\sigma$ is concerned, the Heisenberg principle implies an uncertainty on the tunneling particles momentum and the chiral differential charge current becomes then proportional to the convolution of the system spectral function with a weight function, centered around $k_0$.

The evaluation of the spectral function $\mathcal{A}(eV,k_0)$ for the quenched system \cite{Kennes:2014} is sketched in Appendix \ref{app:spectral} and its features are discussed in the next Section. Here, we want to stress that $\mathcal{G}^Q_\eta	$ in Eq.~\eqref{eq:GQeta} depends only on $\eta$ because of the proportionality factor $A_\eta$. Such a behavior has a remarkable consequence on the charge fractionalization ratio $R^Q$,
\begin{equation}
\label{eq:charge_frac_ratio}
R^Q = \frac{\mathcal{G}_+^Q}{\mathcal{G}_+^Q+\mathcal{G}_-^Q}.
\end{equation}
This ratio represents a convenient tool in order to quantify the amount of charge current which flowed in a given direction \cite{perfetto2014time, calzona2015spin}, let's say the $\eta=+$ direction.   
Using the above definition, from Eq.~\eqref{eq:GQeta} one obtains
\begin{equation}
\label{eq:charge_frac}
R^Q = \frac{A_+}{A_++A_-} = \frac{1+K^f}{2}.
\end{equation}
It is important to note that this is the same expression that one would find for a non-quenched system with interaction strength $K^f$: charge fractionalization has no memory at all of the pre-quench interaction strength $K^i$.
 Furthermore, we note that the ratio $R^Q$, also in this case, has natural bounds, indeed one finds $ 1/2\leq R^{Q}\leq 1 $. 

The physical origin of this behavior and the fact that $R^Q$ is memoryless and insensitive to $K^i$ trace back to the absence of charge transfer between $R$ and $L$ branches. Indeed, the interparticle interaction can only create fluctuations of particle density but the total number of particles on each channel is conserved. Since in our scheme particles are injected solely on the $R$ channel, only this one will actively contribute to the net charge current, no matter how the injection is carried out or how the system is prepared before the injection. As long as only one channel is concerned, it has been shown \cite{perfetto2014time,calzona2015spin} that fractionalization phenomena are completely controlled by the equation of motion of its field operators. Therefore their dynamics depend only on the final (post-quench) Hamiltonian and that is why $R^Q$ does not depend on $K^i$. 

\section{Interaction quench and energy partitioning}\label{sec:energypart}
As mentioned before, interparticle interactions cannot transfer charge between $R$ and $L$ channels. Nevertheless the transfer of energy between the two channels is allowed.
We therefore expect that, in contrast to the behavior of charge fractionalization, the energy flow retains memory of the pre-quench interaction strength $K^i$. In particular, here we will consider the effects of the sudden quench on energy partitioning.
To this end, along the lines of the previous Section, we introduce the chiral energy current in the steady state as 
\begin{equation}
	I_\eta^E = \lim_{t \to \infty} \partial_t\; \int_{-\infty}^{+\infty}\!dx\; \langle \delta \mathcal{H}_\eta (x,t)\rangle,
\end{equation}
identifying the contribution of particles with well-defined energy $eV$ by means of the {\it differential} chiral energy current~\cite{Karzig:2011}
\begin{equation}
	\label{eq:GE_def}
	\mathcal{G}^E_\eta = \frac{\partial I_\eta^E}{\partial V} \Bigg|_{k_P}.
\end{equation}
The explicit calculation of this quantity is reported in Appendix \ref{app:energy} and it yields 
\begin{equation}
	\label{eq:GE}
	\mathcal{G}_\eta^E =  \frac{e}{4 \pi^2 v_F} \int_{-\infty}^{\infty} \!\!dk \; \frac{eV + \eta uk}{2}\, \ \left|\tilde\xi(k-k_0)\right|^2 \,\mathcal{A}(eV,k).
\end{equation}
Not surprisingly, this quantity depends on the spectral function $\mathcal{A}$ and on the weight function $|\tilde \xi|^2$ centered around $k_0$, in analogy with $\mathcal{G}^Q_\eta$ (see Eq. \eqref{eq:GQeta}). There is however a main difference due to the factor $(eV+\eta uk)/2$, related to the energy carried by the excitations.\\
This factor can be understood considering the case of an infinitely extended tunneling ($\sigma \to \infty$), where the single particle energy $eV$ and momentum $k_0$ are conserved by the tunneling process and have to be partitioned between the two counterpropagating chiral excitations in the steady state regime.
These constraints, together with the excitations dispersion relation $\epsilon(k) = \eta u k$, fix the energy of the latter to be $(eV+\eta uk_0)/2$. 

The presence of the $\eta$-dependent factor inside the integral of Eq. \eqref{eq:GE} has a significant impact on the differential chiral currents $ \mathcal{G}^{E}_{\pm} $ and, in particular, on the energy partitioning ratio
\begin{equation}
	\label{eq:RE}
	R^E = \frac{\mathcal{G}_+^E}{\mathcal{G}_+^E+\mathcal{G}_-^E},
\end{equation}
which, as we will see, turns out to be dependent on the pre-quench interaction strength $K^i$. 
Indeed, this ratio depends on the quenched spectral function $ \mathcal{A} $, which qualitatively - and quantitatively - differs from the non-quenched case and it retains memory of the initial state~\cite{Kennes:2014}.
\begin{figure}
	\centering
	\includegraphics[width=\linewidth]{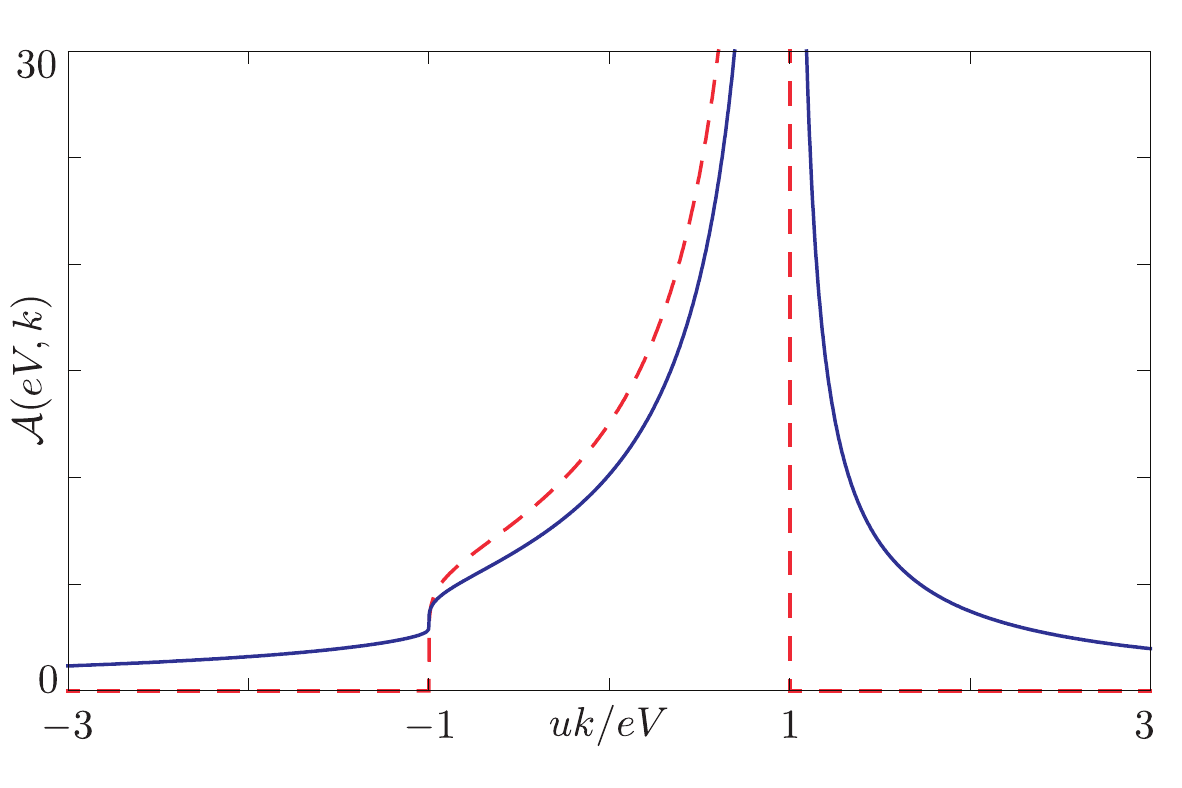}
	\caption{(Color online) Comparison between the spectral function $\mathcal{A}(eV,k)$ (units $ au^{-1} $) of a quenched system (solid blue,  with $K^i = 0.7$ and $K^f = 0.5$) and of a non-quenched one (dashed red, with $K^i=K^f=0.5$). Here $eV a u^{-1} = 5 \cdot10^{-3}$ and $T=0$.} 
	\label{fig:Spectral}
\end{figure}
Figure~\ref{fig:Spectral} shows the quenched spectral function $ \mathcal{A} $ in comparison with the non-quenched one, both at $ T=0 $. These functions are derived in Appendix~\ref{app:spectral} (see Eqs.~\eqref{eq:Aq} and~\eqref{eq:Ath}). The most striking difference between the two is the presence of non-vanishing tails in the quenched case, extending well beyond the range $ -|eV|<uk<|eV| $ (hereafter referred to as inner range). Moreover, it is possible to show that these tails feature a slow power-law decay as $|k|$ increases. This behavior is in sharp contrast with the non-quenched case, in which the spectral function is finite only in the inner range: $\mathcal{A}_{\text{nq}}(eV,k) \propto \theta(|eV|-u|k|)$. 

\begin{figure}
	\centering
	\includegraphics[width=\linewidth]{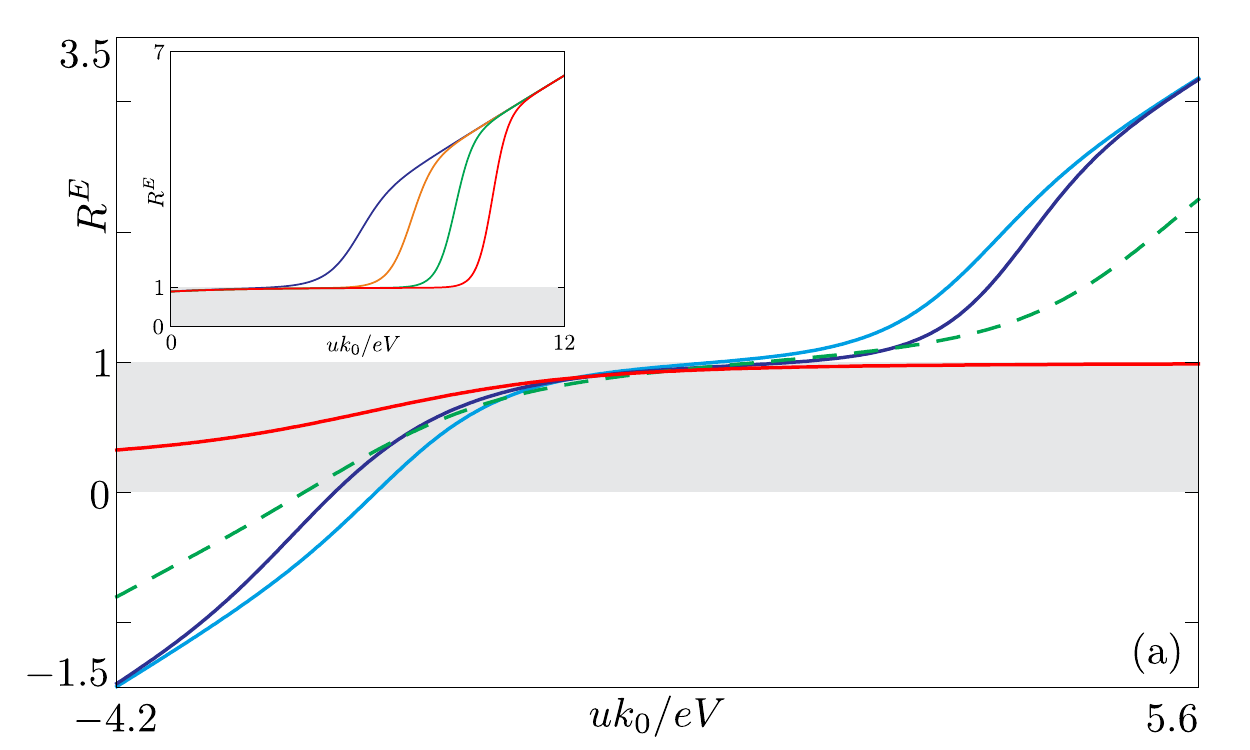}
	\includegraphics[width=\linewidth]{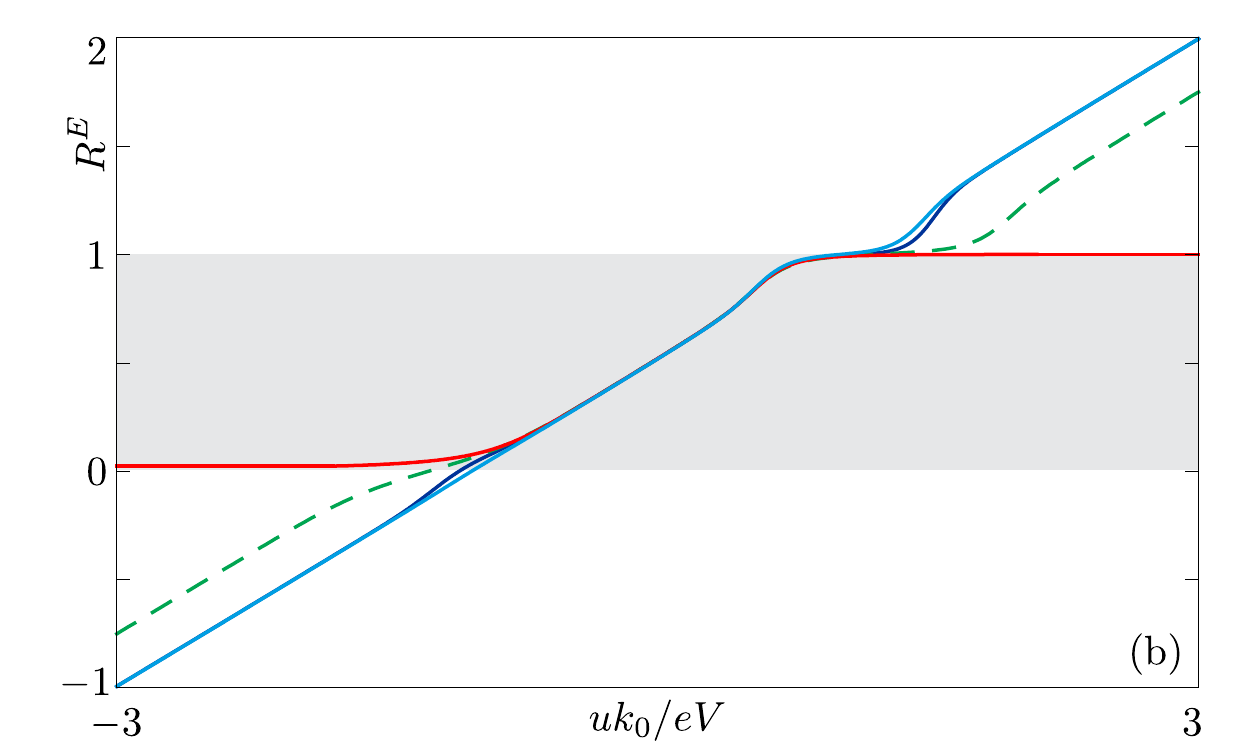}
	\caption{(Color online) Energy partitioning ratio $R^E$ as a function of $uk_0(eV)^{-1}$. Here $K^f = 0.5$ and $eV a u^{-1} = 5\cdot10^{-3}$. In Panel (a) $\sigma eV u^{-1} = 1$ while in Panel (b) $\sigma eV u^{-1} = 10$. In both Panels, the red line depicts the non-quenched case ($K^i=K^f$) whereas the blue and cyan lines refer to quenches from $K^i = 0.7$ and $K^i=1$ respectively. The shaded areas highlight regions with $ 0 \leq R^E \leq 1. $ The dashed green line refers to a non-quenched system at finite temperature $\beta^{-1}=0.02\,  eV$ (see the discussion in the last paragraph of Sec.~\ref{sec:energypart}). The inset in Panel (a) focus on small quenches with $K^f=0.5$ and $\Delta K=10^{-1},10^{-2},10^{-3},10^{-4}$ for the blue, orange, green and red lines respectively.}
	\label{fig:Sigma}
\end{figure}

As shown in Appendix \ref{app:energy} (see Eq.~\eqref{eq:Rnq}), the absence of tails in the non-quenched case necessarily leads to a non-quenched energy partitioning ratio $R_\text{nq}^E$ always constrained by
\begin{equation}
	\label{eq:Re_NQ}
	0 \leq R^E_{\text{nq}} \leq 1.
\end{equation}
On the other hand, in the presence of quench-activated tails the ratio $ R^E $ is not bounded anymore as can be seen in Fig.~\ref{fig:Sigma}.
Here, we compare the quenched energy partitioning ratio (blue and cyan lines) with the non-quenched one (red line) for different quench amplitudes $\Delta K= K^f-K^i$ and for different widths of the tunneling region. The quenched energy partitioning ratio acquires values greater than $1$ (lesser than $0$) when the momentum $k_0$ is sufficiently greater (lesser) than $eV/u$. In these cases, in fact, the Gaussian weight function $|\tilde \xi|^2$ in Eq. \eqref{eq:GE} selects a window well outside the inner range and the presence of quenched-activated tails is thus relevant. Observing the quenched lines in Fig. \ref{fig:Sigma}, the energy partitioning ratio $R^E$ features a linear behavior for sufficiently large $|k_0|$. 
This regime is reached when the contribution to $R^E$ of the tails dominates over the one due to the inner region of the spectral function. Dealing with finite quench amplitudes, like the ones considered in main Panels of Fig. \ref{fig:Sigma}, this regime corresponds to the condition $|uk_0|\gg |eV|+\sigma^{-1}$. In this asymptotic regime we can safely approximate $\mathcal{A}(eV,k)$ by a decaying power-law in Eq. \eqref{eq:GE} and the asymptotic energy partitioning ratio thus reads
\begin{equation}
	R^E \sim \frac{1}{2} + \frac{uk_0}{2eV} \frac{\int_{-\infty}^{+\infty} (1+\frac{k}{k_0}) (1+ \frac{k}{k_0})^{-\zeta} \; e^{-\sigma^2 k^2/2-a|k|}\; dk}{\int_{-\infty}^{+\infty} (1+ \frac{k}{k_0})^{-\zeta} \; e^{-\sigma^2 k^2/2-a|k|}\; dk}
\end{equation}
where $\zeta(K^i,K^f)$ is the exponent of the power-law decay of the tails. Up to first order in $(\sigma k_0)^{-1}$ the energy partitioning factor is $\zeta$-independent and given by the linear function
\begin{equation}
	\label{eq:Re_asym}
	R^E \sim \frac{1}{2} + \frac{u k_0}{2 eV}.
\end{equation}
This result agrees with the main plots in Fig.~\ref{fig:Sigma}. We will comment on the limit $ V\to0 $ later in the discussion. In addition, from the two main Panels (a) and (b) one can easily argue that, fixed all other parameters, the asymptotic regime is reached for smaller momenta when $\sigma$ is increased. 

In general, infinitesimal quenches require more care. The inset shown in Fig. \ref{fig:Sigma}(a) considers the case of very small quenches and shows that the linear behavior in Eq. \eqref{eq:Re_asym} emerges for greater momenta as the quench amplitude $\Delta K$ is reduced. This effect is due to the fact that, as $\Delta K$ decreases, the tails are suppressed and thus their relevance at a fixed $k_0$. In this respect, we note that in the limit of a quench with $\Delta K \to 0$ the tails are infinitesimal and they could be significant only for very large momenta $k_0\to \infty$. As a result, we expect that the energy partitioning factor in presence of an infinitesimal quantum quench will be the same of the non-quenched one for all reasonable values of $k_0$.

\begin{figure}
	\centering
	\includegraphics[width=\linewidth]{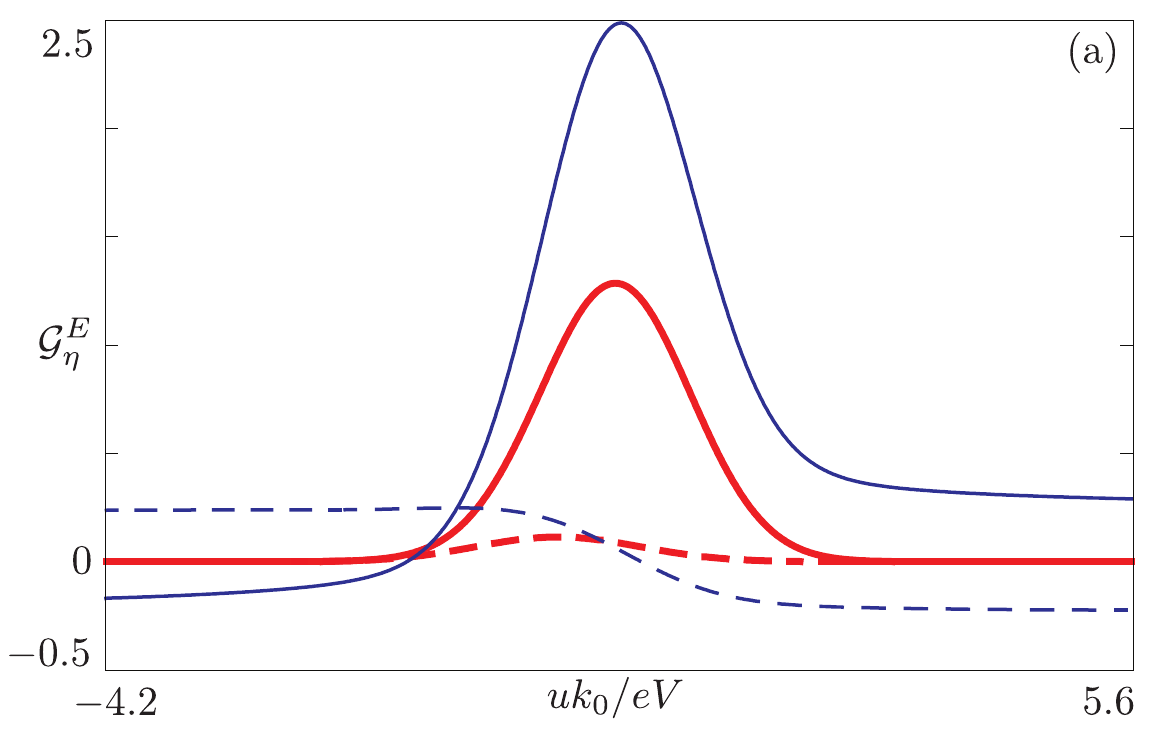}
	\includegraphics[width=\linewidth]{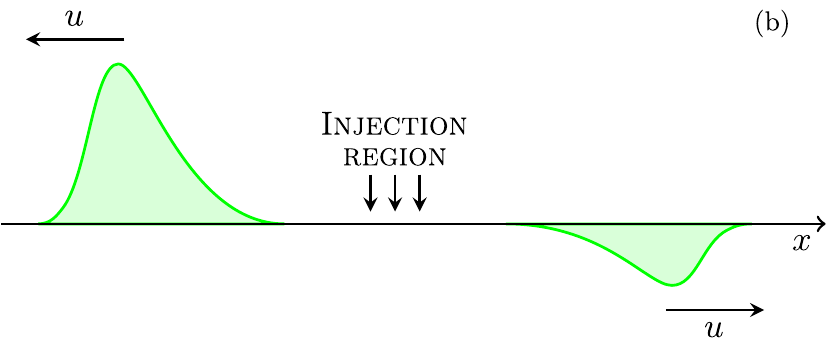}
	\caption{(Color online) Panel (a): Differential energy current $\mathcal{G}_\eta^E$ (units $10^{-3}e|\lambda|^2(4\pi^2 v_F a)^{-1}$) as a function of $uk_0(eV)^{-1}$: comparison between a non-quenched (red thick curves, $K^i=K^f=0.5$) and a quenched (blue curves, $K^i=1$ and $K^f=0.5$) case. Plain curves refers to $\eta=+$ and dashed curves to $\eta=-$. Here $eV a u^{-1} = 5\cdot10^{-3}$ and $\sigma eV u^{-1} = 1$. Panel (b): Cartoon of the contribution to the energy currents of an idealized process of single right-particle injection at energy $ eV>0 $ and momentum $ k_0<0 $. The injected particle produces a negative contribution to the right current and a positive contribution to the left one. } 
	\label{fig:BarP}
\end{figure}

We have shown that a finite quantum quench results in an unbounded $R^E$. This allows for peculiar situations in which the two differential right and left energy currents $\mathcal{G}_\eta^E$  (for $\eta=\pm$) have different sign. Considering for simplicity $eV>0$, this means that the injection of right particles with well-defined energy $eV$ is associated to an energy current traveling to the right with positive or negative sign depending on whether $ k_0\gtrsim0 $ or $ k_0\lesssim0 $ (solid blue line in Fig.~\ref{fig:BarP}(a)) and an energy current traveling to the left with an opposite sign with respect to the right part (blue dashed line in Fig.~\ref{fig:BarP}(a)). In order to further clarify the implications of this fact, we focus on the case with $ k_0<0 $. Here, the injection of right particles with energy $ eV $ produces a positive contribution to the current traveling to the left and a negative contribution to the one traveling to the right (see blue curves in Fig.~\ref{fig:BarP}(a)). This behavior is sketched in Fig.~\ref{fig:BarP}(b) for an idealized case of a single particle injection. Note that the phenomenon of right/left differential energy current with opposite sign is induced by the quench and is completely absent in the non-quenched situation, as can be seen from red curves in Fig.~\ref{fig:BarP}(a). 

It is now worth to analyze the particular case $eV\to 0$ with a finite $k_0$. Here the energy partitioning ratio $R^E$ would diverge (see Eq. \eqref{eq:Re_asym}) meaning that $\mathcal{G}_+^E + \mathcal{G}_-^E=0$  with $ \mathcal{G}^E_{\pm}\neq0 $. Thus, although the tunneling of zero-energy particles does not add energy to the system, as expected, the quench-activated tails of the spectral function allows for a finite $\mathcal{G}_\pm^E$ even with $eV=0$ (see Eq. \eqref{eq:GE}). Interestingly, this means that the tunneling of particles at zero bias in a quenched system results in an energy transfer between the two chiral channels of the system. 

In general, quench-induced effects are relevant also in the non-asymptotic region $-|eV|\lesssim uk_0 \lesssim |eV|$. This can be clearly observed in Fig.~\ref{fig:Sigma}(a), corresponding to a moderate amplitude of the tunneling region ($ \sigma=u(eV)^{-1} $), where in the shaded area there are clear differences among the curves. On the other hand, these differences disappear by increasing the region of tunneling, as can be seen in Fig.~\ref{fig:Sigma}(b), where $ \sigma=10u(eV)^{-1} $. Indeed, in this case the Gaussian weight function $|\tilde \xi|^2$ selects a narrow window peaked around $k_0$. When this region is sufficiently small, the spectral function $ \mathcal{A} $ in Eq.~\eqref{eq:GE} can be approximated as a constant and therefore disappears from the expression of the energy partitioning ratio $ R^E $ in Eq.~\eqref{eq:RE}. Since all the information about the quench is encoded in $ \mathcal{A} $, we thus conclude that in this limit the latter does not depend anymore on the quench amplitude. 

We conclude this Section by commenting on the effects of a finite temperature in a non-quenched LL. The aim is to show that the peculiar features discussed so far are typical of a quench and qualitatively different from the non-quenched thermal case. Indeed, it is well known that also a finite temperature induces tails in the spectral function of a non-quenched LL~\cite{Giuliani:2010,spectralth} (see Eq.~\eqref{eq:Ath}). Their decay is however much faster compared to the one of quenched-activated tails. Thermal tails feature in fact an exponential decay with an associated scale controlled by the temperature dependent factor $(\beta u)^{-1}$, where $ \beta=(k_{\text{B}}T)^{-1} $. Focusing on the asymptotic behavior of $R^E$, i.e. for sufficiently large momenta $|k_0|$ so that the spectral function can be approximated by the contribution of its tails, we obtain
\begin{equation}
	R^E_{\text{th}} \sim \frac{1}{2} + \frac{u}{2 eV} \left[k_0 - \text{sgn}(k_0) \Delta k\right].
\end{equation}
Note that this behavior features the same linear relation we have found for a quenched system in Eq. \eqref{eq:Re_asym} but here a finite shift is present  
\begin{equation}
	\Delta k \propto \frac{u\beta}{\sigma^2}
\end{equation}
which is proportional to the inverse of the temperature. Such a shifted linear behavior emerges clearly in Fig. \ref{fig:Sigma} (green dashed lines). The effects of a finite quantum quench on the energy partitioning ratio are thus qualitatively different from the ones due to a finite (low) temperature in a non-quenched system.

\section{Conclusions}\label{sec:conclusions}
In this work we have studied how peculiar properties of a 1D system, such as charge and energy fractionalization, are modified in the presence of a quench of the interaction strength. We considered two parallel LLs biased with an external dc voltage and tunnel-coupled over a finite size region, one of which is subjected to a sudden interaction quench. We have shown that in the steady state the charge fractionalization ratio retains no memory of the initial state and depends only on the post-quench Hamiltonian. We have ascribed this behavior to charge conservation in each of the LL channels. On the other hand, the energy partitioning is strongly modified by the interaction quench, due to the post-quench reconstruction of spectral function. Indeed, in addition to an overall modification, the quench-activated tails of the latter allow for an unbounded energy partitioning ratio, which corresponds to a situation in which the differential energy current in the two directions have opposite sign. Finally, we have shown that the effects of the quench can be distinguished from thermal non-quenched effects, resulting in different qualitative - and quantitative - behaviors.
The effects discussed in our paper can be probed using a cold atoms setup. In these systems tunneling junctions have recently been realized by optically imprinting a Quantum Point Contact (QPC) at the center of a trapped cloud of fermionic Lithium atoms~\cite{Krinner:2015}. The bias across the QPC can be controlled via connections to particle reservoirs with different particle numbers, yielding a quasi-steady state current and giving direct access to the system transport coefficients. The interaction strength, instead, can be tuned by a magnetic field as done in recent experiments~\cite{Krinner:2015,Krinner:2015b,Husmann:2015}. Breaking the inversion symmetry, required to inject only $ R- $ or $ L- $moving particles, which in a solid state device can be achieved through a magnetic field, is quite subtle in a system of cold atoms since they have a neutral charge. However, it could in principle be implemented through a ``synthetic'' magnetic field, obtainable again by optical means~\cite{Lin:2009}. 
\begin{acknowledgments}
We acknowledge the support of the MIUR-FIRB2013 -- Project Coca (Grant No.~RBFR1379UX) and the COST Action MP1209.
\end{acknowledgments}

\appendix

\section{Time averages}
\label{app:operatorO}
In this Appendix we sketch the derivation of Eq.~\eqref{eq:O_general}, which holds for a generic hermitian and number conserving operator $\hat O$ that acts on the system. That is the case of the operators we are interested in, namely the particle number and the Hamiltonian densities. In the interaction picture with respect to the tunneling Hamiltonian $\hat H_t$ (see Eq.~\eqref{eq:tunneling}), the time evolution average of $\hat O$ reads 
\begin{equation}\langle \hat O(x,t) \rangle = \text{Tr}\left\{\hat O (x,t) \hat \rho(t) \right\},
\end{equation}
with
\begin{equation}
\hat \rho(t) = \mathcal{T}\!\left[e^{-i \int_0^t dt' \hat H_t(t')}\right]\;  \hat \rho(0)\;   \mathcal{\bar T}\!\left[e^{i \int_0^t dt' \hat H_t(t')}\right]
\end{equation}
the density matrix of the whole system. Here, we have assumed that the latter is in thermal equilibrium immediately before the quench (for $t <0$) and $ \hat{\rho}(0) $ is the associated equilibrium density matrix. Furthermore, $ \mathcal{T} $ and $ \mathcal{\bar T} $ denote time-ordering and anti-time-ordering operators respectively. The average variation induced by the tunneling can be thus defined as
\begin{equation}
\langle\delta O (x,t)\rangle = \text{Tr}\left\{\hat O(x,t)  \left[\hat \rho(t) - \hat \rho(0)\right]\right\}.
\end{equation}
At the lowest order in the tunneling one has \cite{calzona2016extended}
\begin{align}
\label{eq:app:deltaO}
\langle\delta O(x,t)\rangle =&\  2 \Re \int_0^{t}\!d\tau_1 \!\int_0^{\tau_1}\!d\tau_2\; \nonumber\\
&  \times \text{Tr}\left\{
\hat \rho(0) \hat H_t^+(\tau_2) \left[\hat O(x,t), \hat H_t^-(\tau_1) \right]
\right.\nonumber\\
&   \left. + \hat \rho(0) \hat H_t^-(\tau_2) \left[\hat O(x,t), \hat H_t^+(\tau_1) \right] \right\},
\end{align}
which directly leads to Eq.~\eqref{eq:O_general}, in view of the fact that $\hat O$ commutes with the probe operators $\hat \chi$ and $\hat \chi^\dagger$.

\section{Charge current}
\label{app:charge}
In order to evaluate the chiral charge current, it is necessary to compute the correlators of the non-interacting biased probe present in Eq.~\eqref{eq:O_general}. We consider a bias protocol such that the probe dispersion relation $\epsilon^p(k) = v_{F} (k_P - k)$ is kept fixed and only its chemical potential is shifted by the bias energy $eV$. Considering $T=0$, one has 
\begin{align}
\langle{\hat \chi_L^\dagger(y_2,\tau_2) \hat \chi_L(y_1,\tau_1)}\rangle =&\ e^{+ i eV (\tau_2-\tau_1)} e^{+ i \left(\frac{eV}{v_F}-k_P \right)(y_2-y_1)}\nonumber\\
&\times f_p(\tau_2-\tau_1,y_2-y_1), \\
\langle{\hat \chi_L(y_2,\tau_2) \hat \chi_L^\dagger(y_1,\tau_1)}\rangle =&\ e^{- i eV (\tau_2-\tau_1)} e^{- i \left(\frac{eV}{v_F}-k_P \right)(y_2-y_1)}\nonumber\\
&\times f_p(\tau_2-\tau_1,y_2-y_1),
\end{align}
with
\begin{equation}
f_p(\tau_2-\tau_1,y_2-y_1) = \frac{1}{2\pi a} \frac{a}{a+i v_F (\tau_2-\tau_1) + i (y_2-y_1)}. 
\end{equation}
Concerning the system part in Eq. \eqref{eq:O_general}, one has to compute
\begin{equation}
\label{eq:commutatore}
\begin{split}
&\left[\hat n_\eta(x,t) , \hat \psi_R^\dagger(y_1,\tau_1)\right] = - \left[\hat n_\eta(x,t) , \hat \psi_R(y_1,\tau_1)\right]^\dagger\\& =\sqrt{K^f} A_\eta \left[\frac{1}{\pi} \frac{a}{a^2+(x-y_1 -u\eta(t-\tau_1))^2}\right] \hat \psi_R^\dagger(y_1,\tau_1).
\end{split}
\end{equation}
The bosonization technique allows to evaluate the interacting correlators which, for $\tau_i>0$, read
\begin{align}
\langle{\hat \psi_R^\dagger(y_2,\tau_2) \hat \psi_R(y_1,\tau_1)}\rangle &= e^{-i k_F (y_2-y_1)} f_s(\tau_1, \tau_2,y_2-y_1),\label{eq:app:c1}\\
\langle{\hat \psi_R(y_2,\tau_2) \hat \psi_R^\dagger(y_1,\tau_1)}\rangle &= e^{i k_F (y_2-y_1)} f_s(\tau_1, \tau_2,y_2-y_1),\label{eq:app:c2}
\end{align}
where 
\begin{equation}
f_s(\tau_1, \tau_2,y_2-y_1) = \frac{1}{2\pi a} \langle e^{-i \sqrt{2\pi} \hat \phi_R(\tau_2, y_2)} e^{i \sqrt{2\pi} \hat \phi_R(\tau_1, y_1)} \rangle.
\end{equation}
The function $f_s$ can be evaluated in two steps. At first one expresses the field $\hat \phi_R$ in terms of the chiral ones $\hat \phi_\eta$, whose time evolution is chiral, i.e. $x-\eta u t$; then such fields are expressed in terms of the pre-quenched ones $\hat \varphi_\eta$, whose average value is known. In the limit $\tau_1,\tau_2 \to \infty$, but keeping $\Delta \tau = \tau_2-\tau_1$ finite, one has
\begin{equation}
\label{eq:app:fsteady}
\begin{split}
f_s(&\tau_1, \tau_2,y) \to f_s^\text{st}(\Delta\tau)\\
=& \frac{1}{2 \pi a} \left(\frac{a}{a+iu \Delta \tau - iy}\right)^{\alpha_+^2A_+^2}
\left(\frac{a}{a+iu \Delta \tau + iy}\right)^{\alpha_+^2A_-^2} \\
& \times \left(\frac{a}{a-iu \Delta \tau + iy}\right)^{\alpha_-^2A_+^2}
\left(\frac{a}{a-iu \Delta \tau - iy}\right)^{\alpha_-^2A_-^2}.
\end{split}
\end{equation}

The calculation now proceed according to Eq. \eqref{eq:IQ}, yielding 
\begin{align}
\label{eq:IQ_app}
I_\eta^Q &=\ \sqrt{K^f}A_\eta\;  2 \Re \int_0^{\infty} \!\!d\tau\! \int_{-\infty}^{+\infty} \!\!\!\!dy_1dy_2\, \xi^*(y_2)\xi(y_1)\nonumber \\
&\times  2 i\;  \sin\left[-\tau eV +(y_2-y_1) \left(\frac{eV}{v_\text{F}} +k_\text{F} - k_P\right)\right]\nonumber\\
&\times f_p(-\tau,y_2-y_1) \, f_s^{\text{st}}(-\tau,y_2-y_1)
\end{align}
Note that, already at this stage of the calculation, it is clear that the charge current depends on $\eta$ only because of the proportionality factor $A_\eta$. As commented in the main text, this immediately proves that charge fractionalization ratio has no memory of the pre-quenched interaction strength $K^i$.

\section{Quenched spectral function and differential charge current}
\label{app:spectral}
Here we compute the quenched spectral function of the $R$-branch of the system, in the steady state limit and at $T=0$. We also derive the expression for the differential chiral charge current given in Eq.~\eqref{eq:GQeta}, where it is presented in terms of the spectral function. The latter is defined as 
\begin{equation}
\mathcal{A}(\epsilon,k) = \int_{-\infty}^{\infty}\!\! d\tau dx\; e^{i (\epsilon \tau - k x)} \lim_{t \to \infty} \mathcal{A}(x,\tau,t),
\end{equation}
where
\begin{equation}
\label{eq:app:A}
\mathcal{A}(x,\tau,t) = \Big\langle \left\{\hat \psi_R(x,t+\tau) , \hat \psi_R^\dagger(0,t)\right\}\Big\rangle 
\end{equation}
and the average is computed with respect to the pre-quench ground state. The R.H.S. of Eq. \eqref{eq:app:A} can be evaluated using the correlators of  Eqs. (\ref{eq:app:c1}),\eqref{eq:app:c2} and \eqref{eq:app:fsteady}. Moreover, it is useful to deal with the functions appearing in $ f_s^{\text{st}} $ in Fourier representation, using the following identity \cite{calzona2016extended}
\begin{equation}
\label{eq:identity}
\left(\frac{a}{a+iz}\right)^g = \frac{(a/u)^g}{\Gamma(g)}\int_{0}^{+\infty} \!\!dE \; E^{g-1} e^{-i E \frac{z}{u}} e^{-E \frac{a}{u}}.
\end{equation}
After some straightforward algebra we get 
\begin{equation}
\label{eq:Aq}
\begin{split}
\mathcal{A}(\epsilon,k) =& \frac{a}{u} \; \frac{\pi}{ \Gamma(A_+^2\alpha_+^2)\Gamma(A_+^2\alpha_-^2)\Gamma(A_-^2\alpha_+^2)\Gamma(A_-^2\alpha_-^2)} \\
&\times\sum_{j=\pm} \prod_{\nu=\pm} \mathcal{I}(A_\nu^2 \alpha_+^2, A_\nu^2 \alpha_-^2, j \Delta_\nu),
\end{split}
\end{equation}
where $\Delta_\pm = (\epsilon \pm uk)/2$ and we have introduced the function\begin{equation}
\begin{split}
\mathcal{I}(C,D,\Delta) &= \int_0^{\infty}\!\!dx\; x^{C-1}\;  \left(x-\frac{a \Delta}{u}\right)^{D-1}\\ &\times \theta\left(x-\frac{a \Delta}{u}\right)\;  e^{\frac{a \Delta}{u}-2x}.
\end{split}
\end{equation}

We can now come back to to the injected chiral current of Eq.~\eqref{eq:IQ_app}, with the aim of expressing it in terms of the quenched spectral function $\mathcal{A}(\epsilon,k)$ in Eq.~\eqref{eq:Aq}. Exploiting again Eq.~\eqref{eq:identity}, $I^Q_\eta$ can be rewritten as 
\begin{equation}
\begin{split}
I^Q_\eta =& \frac{1}{2\pi a u} \; \frac{ \sqrt{K^f}A_\eta}{ \Gamma(A_+^2\alpha_+^2)\Gamma(A_+^2\alpha_-^2)\Gamma(A_-^2\alpha_+^2)\Gamma(A_-^2\alpha_-^2)}\\
&\times \int_0^\infty dEdE_+dE_-dF_+dF_- \;\;  e^{-(E+E_++E_-+F_++F_-)} \\
&\times E_+^{\alpha_+^2 A_+^2} E_-^{\alpha_+^2 A_-^2} F_+^{\alpha_-^2 A_+^2} F_-^{\alpha_-^2 A_-^2} \; \;\\ 
&\times  \sum_{j=\pm 1} j\;  \delta(\tfrac{v_{\rm F}}{u}E+E_++E_--F_+-F_--j\tfrac{a \, eV}{u})\\
& \times \left|\tilde \xi \left(j\left(\tfrac{eV}{v_{\rm F}}+k_{\rm F}-k_P\right) + \tfrac{E_+-E_--F_++F_--E}{a}\right)\right|^2
\end{split}
\end{equation}
Note that the last two lines result from the integration over $\tau$, $y_1$ and $y_2$, see Eq. \eqref{eq:IQ_app}. Here, $\tilde \xi(k)$ is the Fourier transform of the function $\xi(y)$. Taking the derivative with respect to the bias voltage $V$, i.e. focusing on the differential charge current $\mathcal{G}_\eta^Q$, it is possible to identify the expression of the quenched spectral function, evaluated at the bias energy $eV$, and thus obtaining Eq.~\eqref{eq:GQeta}.

We conclude this Appendix by noting that the quenched spectral function present in Eq.~\eqref{eq:Aq} reduces to the non-quenched one
\begin{equation}
\begin{split}
\label{eq:Anq}
\mathcal{A}_\text{nq}(\epsilon,k) =& \frac{2 \pi}{A_-^2\Gamma(A_-^2)^2}\left(\frac{a}{2u}\right)^{2 A_-^2} \; |\epsilon+uk|^{A_-^2}\\
& \times  |\epsilon-uk|^{A_-^2-1}\; \theta(|\epsilon|-u|k|)
\end{split}
\end{equation}
in the limit $\alpha_-\to 0$, i.e. $K^f\to K^i$. Furthermore, in absence of quench but for a finite $\beta=(k_\text{B} T)^{-1}$ and in the limit $a/\beta u \ll 1$, the spectral function is~\cite{Giuliani:2010,spectralth}
\begin{equation}
\label{eq:Ath}
\mathcal{A}_\text{th}(\beta,\epsilon,k) = \frac{\beta}{16 \pi^3}\left(\frac{2\pi a}{\beta u}\right)^{2A_{-}^{2}}\frac{\sum_{\nu=\pm}\mathcal{F}_{\nu}(A_{-},\beta,\epsilon, k)}{\Gamma(A_{-})\Gamma(A_{-}+1)},
\end{equation} 
where
\begin{align}
\mathcal{F}_{\nu}(A,\beta ,\epsilon,k)=& \ e^{\nu\beta\epsilon/2}\left|\Gamma\left[\frac{1}{2}\left(A-\nu \frac{i}{\pi}\beta\Delta_{-}\right)\right]\right|^{2}\nonumber\\
&\times\left|\Gamma\left[\frac{1}{2}\left(A+1+\nu \frac{i}{\pi}\beta\Delta_{+}\right)\right]\right|^{2}.
\end{align}

\section{Differential energy current}
\label{app:energy}
The evaluation of the differential energy current in Eq. \eqref{eq:GE}, although more complicated, follows the same lines of the previous Appendix \ref{app:charge}. In particular, we have to evaluate Eq.~\eqref{eq:O_general} with $\hat O \equiv \mathcal{\hat H}_\eta$. First of all, we calculate the commutator
\begin{equation}
\begin{split}
&\left[\mathcal{\hat H}_\eta(x,t) , \hat \psi_R^\dagger(y_1,\tau_1)\right] = - \left[\mathcal{\hat H}, \hat \psi_R(y_1,\tau_1)\right]^\dagger\\
&\qquad  = \frac{u}{2} \left[\left( \partial_x \hat  \phi_\eta(x-u\eta t)\right)^2\!,\, \hat \psi_R^\dagger(y_1,\tau_1)\right] \\ & \qquad = -\,  \!\frac{u \eta  A_\eta\sqrt{ \pi}}{\sqrt{2}}\!\left[\frac{1}{\pi}\frac{a}{a^2+(x-y_1-u\eta(t-\tau_1))^2}\right] \\
& \qquad\quad  \times \partial_x \left\{ \hat \phi_\eta(x-u\eta t)\,,\, \hat \psi_R^\dagger(y_1,\tau_1) \right\}.
\end{split}
\end{equation}
The structure of this result looks similar to the one in Eq.~\eqref{eq:commutatore} except for the presence of an anticommutator between the fermion field $\hat \psi_R^\dagger(y_1,\tau_1)$ and the chiral boson field $\hat \phi_\eta(x-\eta u t)$. As explained in the main text, it is precisely this $\eta$-dependent anticommutator that originates the energy partitioning dependence on the pre-quench interaction strength $K^i$. In view of Eq. \eqref{eq:O_general}, we have to compute quantum averages like
\begin{equation}
\label{app:mean}
\langle \hat \psi_R^\dagger(y_2,\tau_2) \hat \phi_\eta(z_\eta) \hat \psi_R (y_1,\tau_1) \rangle.
\end{equation}
The explicit evaluation of the quantum average in Eq. \eqref{app:mean} relies again on the bosonization technique and on the identity 
\begin{equation}
\hat \phi_\eta(z) = -i \partial_\nu e^{i \nu \hat \phi_\eta(z)}\Big|_{\nu = 0}.
\end{equation}
The final result for the injected chiral energy current reads
\begin{align}
I_\eta^{E} &=\  \frac{u}{a}\; 2 \Re \int_0^{\infty} \!\!d\tau \int_{-\infty}^{+\infty} \!\!\!\!dy_1dy_2\,\xi^*(y_2)\xi(y_1)\nonumber \\
& \times 2 \cos\left[-\tau eV +(y_2-y_1) \left(\frac{eV}{v_\text{F}} +k_\text{F} - k_P\right)\right]\nonumber\\
& \times f_p(-\tau,y_2-y_1) \, f_s^\text{st}(-\tau,y_2-y_1) \; \mathfrak{F}_\eta(\tau,y_2-y_1).
\end{align}
This result looks like Eq.~\eqref{eq:IQ_app} with the additional $\eta$-dependent factor
\begin{equation}
\mathfrak{F}_\eta(y,\tau) =a A_\eta^2 \left(\frac{\alpha_+^2}{2a-iu\tau -i\eta y} - \frac{\alpha_-^2}{2a+iu\tau +i\eta y}\right). 
\end{equation}
Following the previous Appendix we compute the integrals over $\tau$, $y_1$ and $y_2$ by working in Fourier representation. We thus obtain
\begin{equation}
\begin{split}
I^E_\eta =& \frac{1}{2\pi a u} \; \frac{u}{a} \frac{1}{ \Gamma(A_+^2\alpha_+^2)\Gamma(A_+^2\alpha_-^2)\Gamma(A_-^2\alpha_+^2)\Gamma(A_-^2\alpha_-^2)}\\
&\times \int_0^\infty dEdE_+dE_-dF_+dF_- \;\;  e^{-(E+E_++E_-+F_++F_-)} \\
&\times \sum_{\nu=\pm 1} \nu E_+^{\alpha_+^2 A_+^2} E_-^{\alpha_+^2 A_-^2} F_+^{\alpha_-^2 A_+^2} F_-^{\alpha_-^2 A_-^2} \chi_{\mu,\nu}\; \;\\ 
&\times  \sum_{j=\pm 1} j\;  \delta(\tfrac{v_{\rm F}}{u}E+E_++E_--F_+-F_--j\tfrac{a \, eV}{u})\\
& \times \left|\tilde \xi \left(j\left(\tfrac{eV}{v_{\rm F}}+k_{\rm F}-k_P\right) + \tfrac{E_+-E_--F_++F_--E}{a}\right)\right|^2
\end{split}
\end{equation}
where 
\begin{equation}
\chi_{\mu,\nu} = 
\begin{cases}
E_\mu \quad & \nu=+1\\
F_\mu \quad & \nu=-1
\end{cases}.
\end{equation}
Focusing on the differential energy current $\mathcal{G}_\eta^E$, defined in Eq.~\eqref{eq:GE_def}, it is again possible to identify the expression of the quenched spectral function at the bias energy $eV$ and thus demonstrate the validity of Eq.~\eqref{eq:GE}. Note that it is the adimensional factor $F_\eta(\tau,y)$ which originates the $\eta$-dependent factor $(eV+\eta u k)/2$ appearing in Eq.~\eqref{eq:GE}.

Finally, we comment about the non-quenched case, i.e. when the spectral function satisfies $\mathcal{A}_\text{nq}(eV,k) \propto \theta(|eV|-u|k|)$ (see Eq.~\eqref{eq:Anq}). In that case one has \cite{calzona2016extended,Karzig:2011}
\begin{equation}
\mathcal{G}_+^E = \frac{1}{4 \pi^2 v_F} \int_{-|eV|/u}^{|eV|/u} \!\!dk \; \frac{eV +  uk}{2}\, \ |\tilde\xi(k-k_0)|^2 \,\mathcal{A}_\text{nq}(eV,k).
\end{equation}
Since $\mathcal{A}_\text{nq}(eV,k)\geq0$, one always has the constraint 
\begin{equation}
\begin{split}
(\mathcal{G}_+^E + \mathcal{G}_-^E) \;\leq \;\mathcal{G}_+^E\; \leq \;0 & \qquad eV<0 \\
0\;\leq \;\mathcal{G}_+^E \;\leq \;(\mathcal{G}_+^E + \mathcal{G}_-^E) & \qquad eV\geq 0
\end{split}
\end{equation}
where
\begin{equation}
\mathcal{G}_+^E + \mathcal{G}_- ^E = \frac{1}{4 \pi^2 v_F} \int_{-|eV|/u}^{|eV|/u} \!\!dk \; eV \ |\tilde\xi(k-k_0)|^2 \,\mathcal{A}_\text{nq}(eV,k).
\end{equation}
As a consequence the non-quenched energy partitioning ratio
\begin{equation}
\label{eq:Rnq}
R^E_\text{nq} = \frac{\mathcal{G}_+^E}{\mathcal{G}_+^E+\mathcal{G}_-^E}\end{equation}
is always bounded between $0$ and $1$, as stated in Eq.~\eqref{eq:Re_NQ}.
%\vspace{0.2pt}


\begin{thebibliography}{99}
\bibitem{nota1} Note that this description represents both spinless fermions as well as helical systems.
\bibitem{nota2} Note that we measure charge in units of the particle one so that charge density equals particle density.
\end{thebibliography}
\end{document}